\documentclass{sig-alternate}

\usepackage{mathrsfs}
\usepackage{graphicx}
\usepackage{bbm}
\usepackage{amsmath}
\usepackage{alltt, amssymb}
\usepackage{comment}  
\usepackage{color}
\usepackage{url}            
\usepackage[plainpages=false,pdfpagelabels,colorlinks=true,citecolor=blue,hypert
exnames=false]{hyperref}

\def\bell{\mbox{\boldmath$\ell$}}

\def\blambda{\mbox{\boldmath$\lambda$}}

\newcommand{\bigOsoft}{\tilde{O}}

\def\N {\mathbb{N}}

\def\F {\mathbb{F}}

\def\M{\ensuremath{\mathsf{M}}}
\def\mA {\ensuremath{{\bf A}}}  
\def\mmu {\ensuremath{{\bf u}}}
\def\mv {\ensuremath{{\bf v}}}

\def\mB {\ensuremath{{\bf B}}}

\def\mI {\ensuremath{{\bf I}}}   
\def\mZ {\ensuremath{{\bf 0}}}
\def\mC {\ensuremath{{\bf C}}} 
\def\mD {\ensuremath{{\bf D}}}
\def\mE {\ensuremath{{\bf E}}}
\def\mM {\ensuremath{{\bf M}}}

\def\myproof{\noindent{\sc Proof.}~}
\def\foorp{\hfill$\square$}

\newtheorem{theorem}{Theorem}
\newtheorem{Coro}{Corollary}
\newtheorem{Prop}{Proposition}

\newtheorem{Lemma}{Lemma}

\title{Fast algorithms for differential equations \\ in positive characteristic}

\numberofauthors{2}
\author{
\alignauthor Alin Bostan\\
\affaddr{Algorithms Project}\\
\affaddr{INRIA Rocquencourt}\\ 
\affaddr{France}\\
\affaddr{78153 Le Chesnay Cedex France}\\
\alignauthor \'Eric Schost\\
\affaddr{ORCCA and Computer Science Department}\\
\affaddr{The University of Western Ontario}\\
\affaddr{London, ON, Canada}\\
\affaddr{eschost@uwo.ca} 
}

\begin{document}

\maketitle
\begin{abstract} 
We address complexity issues for linear differential equations in characteristic~$p>0$: resolution and computation of the $p$-curvature. For these tasks, our main focus is on algorithms whose complexity behaves well with respect to~$p$. We prove bounds linear in $p$ on the degree of polynomial solutions and propose algorithms for testing the existence of polynomial solutions in sublinear time $\tilde{O}(p^{1/2})$, and for determining a whole basis of the solution space in quasi-linear time $\tilde{O}(p)$; the $\tilde{O}$ notation indicates that we hide logarithmic factors. We show that for equations of arbitrary order, the $p$-curvature can be computed in subquadratic time $\tilde{O}(p^{1.79})$, and that this can be improved to $O(\log(p))$ for first order equations and to $\tilde{O}(p)$ for classes of second order equations.  
\end{abstract}


\vspace{1mm}
 \noindent
 {\bf Categories and Subject Descriptors:} \\
\noindent I.1.2 [{\bf Computing Methodologies}]:{~} Symbolic and Algebraic
  Manipulation -- \emph{Algebraic Algorithms}
 
 \vspace{1mm}
 \noindent
 {\bf General Terms:} Algorithms, Theory
 
 \vspace{1mm}


\medskip

\section{Introduction}\label{sec:intro}

\noindent We study several algorithmic questions related to linear
differential equations in characteristic~$p$, where $p$ is a prime
number: resolution of such equations and computation of their
$p$-curvature. Our emphasis is on the complexity viewpoint.

Let thus $\F_p$ be the finite field with $p$ elements, and let $\F_p(x)\langle
\partial \rangle$ be the algebra of differential operators with
coefficients in $\F_p(x)$, with the commutation relation $\partial x =
x\partial + 1$. One of the important objects associated to a
differential operator $L$ of order $r$ in $\F_p(x)\langle \partial
\rangle$ is its $p$-curvature, hereafter denoted $\mA_p$. By
definition, this is the $(r\times r)$ matrix with coefficients in
$\F_p(x)$, whose $(i,j)$-entry is the coefficient of $\partial^{i}$ in
the remainder of the Euclidean (right) division of $\partial^{p+j}$ by
$L$, for $0 \le i,j < r$.

The concept of $p$-curvature originates in Grothendieck's work in the
late 1960s, in connection to one of his famous (still unsolved)
conjectures. In its simplest form, this conjecture is an arithmetic
criterion of algebraicity, which states that a linear differential
equation with coefficients in $\mathbb{Q}(x)$ has a basis of algebraic
solutions over $\mathbb{Q}(x)$ if and only if its reductions
modulo~$p$ have zero $p$-curvature, for almost all primes~$p$. The
search of a proof of this criterion motivated the development of a
theory of differential equations in characteristic~$p$ by
Katz~\cite{Katz70}, Dwork~\cite{Dwork82}, Honda~\cite{Honda81}, etc.

There are two basic differences between differential equations in
characteristic zero and~$p$: one concerns the dimension of the
solution space, the other, the form of the solutions. While in
characteristic zero, a linear differential equation of order $r$
admits exactly $r$ linearly independent solutions, this is no longer
true in positive characteristic: for $L\in \F_p(x)\langle \partial
\rangle$, the dimension of the solution space of the equation $Ly=0$
over the field of constants $\F_p(x^p)$ is generally less than the
order~$r$. Moreover, by a theorem of Cartier and Katz (see
Lemma~\ref{lemma:pcurv-vs-ratsols} below), the dimension is exactly
$r$ if and only if the $p$-curvature matrix $\mA_p$ is zero. Thus,
roughly speaking, the $p$-curvature measures to what extent the
solution space of a differential equation modulo~$p$ has dimension
close to its order. 

On the other hand, the form of the solutions is simpler in
characteristic~$p$ than in characteristic zero. Precisely, the
existence of polynomial solutions is equivalent to the existence of
solutions which are either algebraic over $\F_p(x)$, or power series
in $\F_p[[x]]$, or rational functions in $\F_p(x)$~\cite{Honda81}.
Therefore, in what follows, by solving $Ly=0$ we simply understand
finding its polynomial solutions.

In computer algebra, the $p$-curvature was publicised by van der
Put~\cite{vanDerPut95,vanDerPut96}, who used it as a central tool in
designing algorithms for factoring differential operators in
$\F_p(x)\langle \partial \rangle$. Recently, his algorithms were
analyzed from the complexity perspective and implemented by
Cluzeau~\cite{Cluzeau03}, who extended them to the case of
systems. Cluzeau also took in~\cite{Cluzeau04} a first step towards a
systematic modular approach to the algorithmic treatment of
differential equations.

Improving the complexity of the $p$-curvature computation is an
interesting problem in its own right. Our main motivation for studying
this question comes, however, from concrete applications. First, in a
combinatorial context, the use of the $p$-curvature served in the
automatic classification of restricted lattice walks~\cite{BoKa08b}
and notably provided crucial help in the treatment of the notoriously
difficult case of Gessel's walks~\cite{BoKa08a}.  Also, intensive
$p$-curvature computations were needed in~\cite{BBHMWZ08}, where the
question is to decide whether various differential operators arising
in statistical physics have nilpotent, or zero, $p$-curvature. 

In the latter questions, the prime $p$ was ``large'', typically of the
order of $10^4$. This remark motivates our choice of considering $p$
as the most important parameter: our primary objective is to obtain
complexity estimates featuring a low exponent in $p$.

\smallskip\noindent{\bf Previous work.}  The non-commutativity of
$\F_p(x)\langle \partial \rangle$ prevents one from straightforwardly
using binary powering techniques for the computation of $\mA_p$ via
that of $\partial^p \bmod L$. Thus, the complexity of all currently
known algorithms for computing the $p$-curvature is quadratic in~$p$.

Katz~\cite{Katz82} gave the first algorithm, based on the following
matrix recurrence: define
\begin{equation}
  \label{eq:pcurv}
\mA_1 = \mA,\quad \mA_{k+1} = \mA_k' + \mA \mA_k,
\end{equation}
where $\mA \in \mathscr{M}_r(\F_p(x))$ is the companion matrix
associated to $L$; then, $\mA_p$ is the $p$-curvature matrix (hence
our notation). 

It was observed in~\cite[\S13.2.2]{PuSi03} that it is slightly more
efficient to replace~\eqref{eq:pcurv} by the recurrence $\mv_{k+1} =
\mv_k' + \mA \mv_k$ which computes the first column $\mv_k$ of
$\mA_k$, by taking for $\mv_0$ the first column of $\mI_r$. Then
$\mv_p,\ldots,\mv_{p+r-1}$ are the columns of~$\mA_p$. This
alternative requires only matrix-vector products, and thus saves a
factor of~$r$, but still remains quadratic in~$p$.  Cluzeau proposed
in~\cite[Prop.~3.2]{Cluzeau03} a fraction-free version
of~\eqref{eq:pcurv} having essentially the same complexity, but
incorrectly stated that the method in~\cite{PuSi03} works in linear
time in~$p$.

Concerning polynomial and rational solutions of differential equations
modulo~$p$, very few algorithms can be found in the literature. Cluzeau
proposes in~\cite[\S2]{Cluzeau03} an algorithm of cubic complexity
in~$p$ and, in the special case when $\mA_p=0$, a different algorithm
of quadratic complexity in~$p$, based on a formula due to Katz which is
the nub of Lemma~\ref{lemma:pcurv-vs-ratsols} below.

\smallskip\noindent{\bf Our contribution.}  We prove in
Section~\ref{sec:solpol} a linear bound in $p$ on the degree for a
basis of the solution space of polynomial solutions of an equation
$Ly=0$. Then, we adapt the algorithm in~\cite{AbBrPe95} and its
improvements~\cite{BoClSa05} to the case of positive characteristic;
we show how to test the existence of polynomial solutions in time
nearly proportional to~$p^{1/2}$, and how to determine a full basis of the solution
space in time quasi-linear in~$p$.

Regarding the $p$-curvature, we first focus on two particular cases:
first order operators, where the cost is polynomial in $\log(p)$
(Section~\ref{sec:one}), and second order ones, for which we obtain a
cost quasi-linear in~$p$ in some cases 
(Section~\ref{sec:two}).

In general, a useful way to see~\eqref{eq:pcurv} is to note that the
$p$-curvature is obtained by applying the operator
$(\partial+\mA)^{p-1}$ to~$\mA$. In Section~\ref{sec:high} we exploit
this observation. As a side result, we give a baby steps~/~giant steps
algorithm for computing the image $Lu$ of an operator $L$ applied to a
polynomial $u$; this is inspired by Brent-Kung's algorithm for power
series composition~\cite{BrKu78}.

\smallskip\noindent{\bf Complexity measures.} Time complexities are
measured in terms of arithmetic operations in~$\F_p$. 

We let $\M: \N \rightarrow \N$ be such that polynomials of degree at
most~$n$ in~$\F_p[x]$ can be multiplied in time $\M(n)$. Furthermore,
we assume that $\M(n)$ satisfies the usual assumptions
of~\cite[\S8.3]{GaGe99}; using Fast Fourier Transform, $\M(n)$ can be
taken in $O(n \log n\,\log\log n)$~\cite{ScSt71,CaKa91}.  We suppose
that $2\leq \omega \le 3$ is a constant such that two matrices in
$\mathscr{M}_n(\F_p)$ can be multiplied in time $O(n^\omega)$. The
current tightest upper bound is $\omega < 2.376$~\cite{CoWi90}.

The precise complexity estimates of our algorithms are sometimes
quite complex; to highly their main features, we rather give
simplified estimates. Thus, we use the notation $f \in \bigOsoft(g)$
for $f,g:\N \rightarrow \N$ if $f$ is in $O(g \log(g)^m)$ for some
$m\geq 1$. For instance, $\M(n)$ is in $\bigOsoft(n)$.


\section{Preliminaries}

\noindent {\bf Basic properties of the $p$-curvature.} 
We first give degree bounds on the $p$-curvature of an operator.
Consider 
\begin{equation}
  \label{eq:L}
L = \ell_0(x) + \ell_1(x) \partial + \cdots + \ell_r(x) \partial^r,  
\end{equation}
with all $\ell_i$ in $\F_p[x]$ of degrees at most $d$ and $\ell_r \ne
0$. As in~\eqref{eq:pcurv}, we define $\mA_1 = \mA$ and $\mA_{k+1} =
\mA_k' + \mA \mA_k$ for $k \ge 1$.
\begin{Lemma}\label{lemma:1}
  For $k \ge 0$, let $\mB_k = \ell_r^k \mA_k$. Then $\mB_k$ is in
  $\mathscr{M}_r(\F_p[x])$, with entries of degree at most $dk$.
\end{Lemma}
\myproof
Explicitly, we have
$$\mA = \left [ \begin{matrix}
  & &&  -\frac{\ell_0}{\ell_r} \\
1 & &&   -\frac{\ell_1}{\ell_r} \\
  &\ddots &&  \vdots \\
  &     &1& -\frac{\ell_{r-1}}{\ell_r}
\end{matrix} \right ].$$ From this, we see that the sequence $\mB_k$
satisfies the equation
$$\mB_{k+1} =  \ell_r \mB_k' + (\mB_1 - k \ell'_r \mI_r)\mB_k,$$
where $\mI_r$ is the $r\times r$ identity matrix. The claim
follows.  \foorp

\smallskip\noindent In particular, the $p$-curvature $\mA_p$ has
the form $\mB_p/\ell_r^p$, with $\mB_p$ a polynomial matrix of degree
at most $dp$.


A second useful result is the following lemma, attributed to Katz. It
relates the solution space of $Ly=0$ to the $p$-curvature and
generalizes a theorem of Cartier.  A proof can be found
in~\cite[Th.~3.8]{Cluzeau03}.


\begin{Lemma}\label{lemma:pcurv-vs-ratsols}
  The dimension over $\F_p(x^p)$ of the vector space of rational
  solutions of $L$ is equal to the dimension over $\F_p(x)$ of the
  kernel of $\mA_p$.  In particular, $L$ has a basis of polynomial
  solutions if and only if its $p$-curvature is zero.
\end{Lemma}

\smallskip\noindent{\bf Operator algebras.} In what follows, we mainly
consider operators with coefficients in $\F_p(x)$, but also sometimes
more generally in the $(n\times n)$ matrix algebra
$\mathscr{M}_n(\F_p(x))$;  as has been done up to now, we will write 
matrices in bold face. If $L$ is in
$\mathscr{M}_n(\F_p(x))\langle \partial \rangle$ of the form
$$L = \bell_0(x) + \bell_1(x) \partial + \cdots + \bell_r(x) \partial^r,$$
with coefficient matrices $\bell_i$ in $\mathscr{M}_n(\F_p[x])$ of
maximal degree $d$, we say that $L$ has {\em bidegree} $(d,r)$. 

\smallskip\noindent{\bf Regularization.} For most of our algorithms,
we must assume that the origin $x=0$ does not cancel the leading term
$\ell_r \in \F_p[x]$ of the operator $L$. 

If we can find $x' \in \F_p$ such that $\ell_r(x') \ne 0$, we can
ensure this property by translating the origin to $x'$. To ensure that
we can find $x'$, we must make the following hypothesis, written ${\bf
  H}$: $\ell_r$ does not vanish identically on $\F_p$.
\begin{Lemma}
  Given $L$ of bidegree $(d,r)$, testing whether ${\bf H}$ holds
  can be done in time $O(\M(d)) \subset \bigOsoft(d)$. If so, one can find $x'$
  such that $\ell_r(x') \ne 0$ and translate the coordinates' origin
  to $x'$ in time $O(r\M(d)\log(d)) \subset \bigOsoft(rd)$.
\end{Lemma}
\myproof Testing ${\bf H}$ amounts to verify whether $x^p-x$ divides
$\ell_r$. If $\deg(\ell_r) < p$, $\bf H$ obviously holds. Else, we
have $p \le d$; then, it is enough to reduce $\ell_r$ modulo $x^p-x$, which
takes time $O(\M(d))$.

If $\bf H$ holds, we know that we can find $x' \in \{0,\dots,\deg(\ell_r)\}$ such that $\ell_r(x')
\ne 0$; so it is enough to evaluate
$\ell_r$ at this set of points, which by~\cite[\S 10.1]{GaGe99} takes time
$O(\M(d)\log(d))$. Once $x'$ is known, we shift all coefficients of
$L$ by $x'$. Using fast algorithms for polynomial shift~\cite{GaGe97},
the time is $O(\M(d)\log(d))$ per coefficient; the conclusion follows.
\foorp

\smallskip\noindent As a consequence, in all the following algorithms, we
will assume that ${\bf H}$ holds. If not, one could actually work in a
low-degree extension of $\F_p$ to find $x'$; we do not consider this 
generalization here.


\section{Polynomial solutions}\label{sec:solpol}

\noindent We start with the study of the polynomial solutions of a
linear differential equation; aside from its own interest, this
question will arise in our algorithm for order two operators in
Section~\ref{sec:two}. 
\begin{theorem}\label{coro:solpol}
  Let $L$ be as in~\eqref{eq:L}, with $r \le d$ and $r \le p$, and such
  that ${\bf H}$ holds. Then, one can test whether the equation $Lu=0$
  has non-zero solutions in $\F_p(x)$ in time
  $$\bigOsoft(d^\omega r^{1/2} p^{1/2}+ d^{\omega+1}r^{\omega-1}).$$ 
  If so, one can determine a basis of the solution set consisting of
  polynomials of degree at most $dp-1$ in extra time
$$\bigOsoft(d^{\omega+1}rp + d^2 r^{\omega+3} p).$$
\end{theorem}
\noindent The main point here is that for fixed $d$ and $r$, {\em
  testing} the existence of solutions takes time
$\bigOsoft(p^{1/2})$, whereas finding a basis of the solution space
takes time $\bigOsoft(p)$.

In all this section, $L$ is fixed, and the assumptions of
Theorem~\ref{coro:solpol} are satisfied. The assumptions on the
relative order of magnitude of $p,d,r$ help us obtain simple cost
estimates and rule out some possible overlaps in indices modulo
$p$. The assumption $r \le d$ is here mostly for convenience; the
assumption $r\le p$ is necessary.


\subsection{Degree bounds}

\noindent 
Let $\mathcal{F}$ be the $\F_p(x^p)$-vector space of rational
solutions of the equation $L u=0$.  The following proposition proves a
bound linear in $p$ on the degree of a basis of $\mathcal{F}$. To our
knowledge, such linear bounds were previously available only in two
particular cases: (a) when the equation has a basis of polynomial
solutions and under the additional hypotheses $0\le \deg(\ell_0) - r
\le p - 1$ and $p \ge r$~\cite[Th.~7]{Honda81}; (b) when $r = 2$ and
the equation has exactly one nonzero polynomial
solution~\cite[Lemma~10.1]{Dwork82}. These bounds are respectively
$(p-r)d+ {r\choose 2}$ for (a) and $\frac12 (p-1)(d-1)$ for (b). In the
general case, the analysis in~\cite{Cluzeau03,Cluzeau04} suggests a
bound quadratic in $p$ of type $p(p + d)$. Our result refines this
approach.

\begin{Prop}\label{theo:theorysolpol}
  If $Lu = 0$ has at least one nonzero solution in $\F_p(x)$, then
  $\mathcal{F}$ admits a basis consisting of polynomial solutions of
  degree at most $pd-1$ each.
\end{Prop}
\myproof The map $\varphi_L: \F_p(x) \to \F_p(x)$ defined by $y
\mapsto L(y)$ is $\F_p(x^p)$-linear. Let $\mM \in
\mathscr{M}_p(\F_p(x^p))$ be the matrix of this map with respect to the basis
$(1, x,\dots, x^{p-1})$.  Write $\mM = (m_{i,j})_{0\le i,j\le
  p-1}$ for some $m_{i,j}$ in $\F_p[x^p]$. Then, $u \in \F_p[x]$ is in
$\mathcal{F}$ if and only if $\mM \times [u_0 \cdots u_{p-1}]^t = 0$,
with $u_i$ in $\F_p[x^p]$ such that $u = u_0 +  u_1x + \dots +
u_{p-1}x^{p-1}$.

Since $L(x^i) = \sum_{j\le p-1} m_{i,j}x^j$ is a sum of $p$
polynomials of pairwise distinct degrees $\deg(m_{i,j})+j$, we deduce
that for all $i,j$, $\deg(m_{i,j})+j \le \deg(L(x^i))$.

Since $Lu=0$ has a non-zero solution in $\F_p(x)$, it has also a
non-zero solution in $\F_p[x]$, by clearing denominators. Let thus $v$
be in $\F_p[x]\setminus\{0\}$ such that $Lv=0$, or equivalently $\ell_0 v =
-\sum_{1 \le j \le r} \ell_j v^{(j)}$. Since all terms in the
right-hand side have degree at most $d+\deg(v)-1$, we deduce that
$\deg(\ell_0) \le d-1$. This implies that $L(x^i)=\ell_0 x^i + 
\sum_{1 \le j \le r} i\cdots(i-j+1) \ell_i x^{i-j}$ has degree
at most $d+i-1$.

To summarize, for all $0 \le i,j \le p-1$, we obtain the inequality
$\deg(m_{i,j}) \le (d-1) + (i-j)$. This implies that for 
any permutation $\sigma$ of $\{0,\dots,p-1\}$, 
$$
\begin{array}{c}
  \deg (\prod_{i=0}^{p-1} m_{i,\sigma(i)})
  =\sum_{i=0}^{p-1} \deg(m_{i,\sigma(i)})
  \le p(d-1),\end{array}$$
since the sum of the terms $i-\sigma(i)$ is zero. This implies
that all minors of $\mM$ have degree at most $p(d-1)$, since
any term appearing in the expansion of such minors can be completed
to form one of the form $\prod_{0 \le i \le p-1} m_{i,\sigma(i)}$.

The nullspace of $\mM$ admits a basis $[\mv_1,\dots,\mv_k]$, all of whose 
entries are minors of $\mM$. By what was said above, they all have
degree at most $p(d-1)$. A basis of $\mathcal{F}$ is easily deduced:
to $\mv_i=[v_{i,0}\cdots v_{i,p-1}]^t$ corresponds the polynomial
$v_i = v_{i,0} + \dots + v_{i,p-1}x^{p-1}$. We deduce that
$\deg(v_i) \le p-1+p(d-1)=pd-1$, as claimed. \foorp


\subsection{Solutions of bounded degree}
 
\noindent Let $\mathcal{G}\subset\F_p[x]$ be the $\F_p$-vector space
of polynomial solutions of $Lu=0$ of degree at most $pd-1$. We are
interested in computing either the dimension of $\mathcal{G}$, or an
$\F_p$-basis of it. In view of the former proposition, this will be
sufficient to prove
Theorem~\ref{coro:solpol}. Proposition~\ref{theo:solpol} gives cost
estimates for these tasks, adapting the algorithm in~\cite{AbBrPe95}
and its improvements~\cite{BoClSa05}.
\begin{Prop}\label{theo:solpol}
  Under the assumptions of Theorem~\ref{coro:solpol}, one can compute
  $\dim_{\F_p}(\mathcal{G})$ in time
  $$\bigOsoft(d^\omega r^{1/2} p^{1/2}+ d^{\omega+1}r^{\omega-1}).$$
One can deduce a basis of $\mathcal{G}$ in extra time
$\bigOsoft(d^{\omega+1}rp)$.
\end{Prop}
For $r$ and $d$ fixed, the main feature of this result is that the
cost of computing the dimension of $\mathcal{G}$ is the sublinear
$\bigOsoft(p^{1/2})$, whereas the cost of computing a basis of it is
$\bigOsoft(p)$.

\smallskip \myproof Let $u_0,\dots,u_{pd-1}$ be unknowns and let
$u$ be the polynomial
$u=\sum_{n < pd} u_n x^n$; for $n<0$ or $n\ge pd$, we let
$u_n=0$. There exist $c_0,\dots,c_{d+r}$ in $\F_p[n]$, of degree at
most $r$, such that for $n \ge 0$, the coefficient of degree $n$ of
\begin{equation}
  \label{eq:u}
\ell_0(x) u + \cdots + \ell_r(x) u^{(r)}
\end{equation}
is 
$C_n= c_0(n) u_{n-d}+ \cdots + c_{d+r}(n) u_{n+r};$
note for further use that
\begin{equation}
  \label{eq:coef}
C_{n-r}= c_0(n-r) u_{n-r-d}+ \cdots + c_{d+r}(n-r) u_n.
\end{equation}
The polynomial $u$ is in $\mathcal{G}$ if and only if $C_n=0$ for $0
\le n \le (p+1)d-1$. Shifting indices, we obtain the system of linear
equations $C_{n-r}=0$, with $r \le n \le (p+1)d+r-1$, in the unknowns
$u_0,\dots,u_{pd-1}$.

The matrix of this system is band-diagonal, with a band of width
$d+r+1$. In characteristic zero or large enough, one can eliminate each
unknown $u_n$, with $n \ge r$, using $C_{n-r}$. Here, some equations
$C_{n-r}$ become deficient, in the sense that the coefficient of
$u_n$ vanishes; this induces a few complications.

\smallskip\noindent{\bf Outline of the computation.}
Since $\lambda=\ell_r(0)$ is not zero, $c_{d+r}(n)$ is the non-zero
polynomial $\lambda(n+1)\cdots(n+r)$, and
$c_{d+r}(n-r)=\lambda(n-(r-1))\cdots n$.  Let then $R=[0,\dots,r-1]$
be the set of roots of the latter polynomial. For $r \le n \le pd-1$,
if $(n \bmod p)$ is not in $R$, then $u_n$ is the highest-index
unknown appearing with a non-zero coefficient in $C_{n-r}$; we can
then eliminate it, by expressing it in terms of the previous $u_m$'s.

The unknowns we cannot eliminate this way are $u_n$, with $n$ in 
$$A=[\ n \ |\ 0 \le n \le pd-1,\ (n \bmod p) \in R\ ];$$ 
the residual equations are $C_{n-r}=0$,
for $n$ in $B=B_1 \cup B_2$, with
$$B_1=[\ n \ | \ r \le n \le pd-1 \text{~and~} (n\bmod p) \in R\ ]$$
and 
$$B_2=[\ n \ | \ pd \le n \le (p+1)d+r-1\ ].$$
To determine the dimension of $\mathcal{G}$, and later on find a basis
of it, we rewrite the residual equations using the residual unknowns.

For $n=ip+j$ in $B_1$, the unknowns present in $C_{n-r}$ are
$u_{ip+j-r-d},\dots,u_{ip+j}$. Of those, only
$u_{ip+j-r-d},\dots,u_{ip-1}$ need to be rewritten in terms of $[u_n\
|\ n \in A]$; the others already belong to this set. Thus, it is
enough to express all $u_{ip-r-d},\dots,u_{ip-1}$ in terms of $[u_n\
|\ n \in A]$, for $1 \le i < d$.

For $n$ in $B_2$, the unknowns in $C_{n-r}$ are
$u_{n-r-d},\dots,u_{pd-1}$ (the higher index ones are zero). So, it is
enough to compute $u_{pd-r-d},\dots,u_{pd-1}$ in terms of $[u_n\ |\
n \in A]$. This is thus the same problem as above, for index $i=d$.

\smallskip\noindent{\bf Expressing all needed unknowns using $A$.}  Let
$A'=[0,\dots,pd-1]-A$. For $n$ in $A'$, one can rewrite the equation
$C_{n-r}=0$ as the first order recurrence
\begin{equation}
  \label{eq:rec}
\left [ \begin{matrix}
u_{n-r-d+1}\\[-1mm]
\vdots \\
u_{n}
\end{matrix} \right ] = 
\mA(n) \left [ \begin{matrix}
u_{n-d-r}\\[-1mm]
\vdots \\
u_{n-1}
\end{matrix} \right ]  
\end{equation}
with
$$
\mA(n)=\left [ \begin{matrix}
0&1&\dots&0\\
0&0&\ddots&0\\
0&0&\dots&1\\
-\frac{c_{0}(n-r)}{c_{d+r}(n-r)} &\dots&\dots& -\frac{c_{d+r-1}(n-r)}{c_{d+r}(n-r)}
\end{matrix} \right ];$$ note that for $n\ne 0\bmod p$,
$\mA(n+p)=\mA(n)$. Let next $\mB$ be the matrix factorial
$\mA(p-1)\cdots\mA(r)$. Then we have the equalities, for $1 \le i \le d$:
$$
\left [ \begin{matrix}
u_{ip-r-d}\\[-1mm]
\vdots \\
u_{ip-1}
\end{matrix} \right ] = 
\mB \left [ \begin{matrix}
u_{(i-1)p-d}\\[-1mm]
\vdots \\
u_{(i-1)p+r-1}
\end{matrix} \right ].
$$
Note that $|A|=dr$; we let $\mmu$ be the $dr\times 1$
column-vector consisting of all $u_n$, for $n$ in $A$. Let further
$\mC_0$ be the $(d+r)\times dr$ zero matrix. For $1 \le i \le d$,
suppose that we have determined $(d+r)\times dr$ matrices 
$\mC_1,\dots,\mC_{i-1}$ such that, for $1 \le j < i$, we have
\begin{equation}
  \label{eq:Cj}
\left [ \begin{matrix}
u_{jp-r-d}\\[-1mm]
\vdots \\
u_{jp-1}
\end{matrix} \right ] = 
\mC_j \mmu
\quad\text{and}\quad
\left [ \begin{matrix}
u_{jp-r-d}\\[-1mm]
\vdots \\
u_{jp+r-1}
\end{matrix} \right ] = \mD_j \mmu,
\end{equation}
with
$$
\mD_j=\left [ \begin{matrix}   \mC_j   \\ \mZ_{r\times \ell_j}~\mI_r~\mZ_{r \times \ell'_j}
  \end{matrix} \right ], \;\ell_j = (j-1)r\quad\text{and}\; \ell'_j=(d-j-1)r.$$  
Letting $\mC'_{i-1}$ be the matrix made of the last $d$ rows of
$\mC_{i-1}$, we define
$$\mC_i = \mB \left [ \begin{matrix}   \mC_{i-1}'  \\ \mZ_{r\times \ell_{i-1}}~\mI_r~\mZ_{r \times \ell'_{i-1}} \end{matrix} \right ];$$ then,~\eqref{eq:Cj} is satisfied at index
$i$ as well. 

\smallskip\noindent{\bf Rewriting all residual equations using $A$.}
Combining all previous information, we obtain a matrix equality of the
form $\mmu' = \mD \mmu$, where $\mmu'$ is the column vector with entries
$u_{ip-r-d},\dots,u_{ip+r-1}$, for $1 \le i \le d$, and where $\mD$ is
the matrix obtained by stacking up $\mD_1,\dots,\mD_d$.

We have seen that all indeterminates appearing in the residual
equations $C_{n-r}$, with $r$ in $B$, are actually in $\mmu'$.  By
evaluating the coefficients $c_0,\dots,c_{d+r}$ at $n-r$, for $n$ in
$B$, we obtain the matrix $\mD'$ of the residual equations, expressed
in terms of the unknowns in $\mmu'$. Hence, the matrix $\mE=\mD'\mD$
expresses the residual equations in terms of $u_n$, for $n$ in $A$.

By construction, the dimension of $\mathcal{G}$ equals the dimension
of the nullspace of $\mE$. Knowing a basis of the nullspace of $\mE$,
one deduces a basis of $\mathcal{G}$ using~\eqref{eq:rec}, to compute
all $u_n$ for $n$ in $A'$.

\smallskip\noindent{\bf Cost analysis.} By~\cite[Lemma~7]{BoClSa05},
one can compute~$\mB$ in time $T_1 \!=\! O(d^\omega\M(r^{1/2}p^{1/2}) \log(rp))$,
which is $\bigOsoft(d^\omega r^{1/2} p^{1/2})$.
Computing a matrix $\mC_i$ requires one matrix multiplication of size
$(d+r,d) \times (d,dr)$. In view of the inequality $r \le d$,
using block matrix multiplication, this can be done in time
$O(d^\omega r)$. Thus, computing all needed matrices $\mC_i$ takes time
$T_2=O(d^{\omega+1} r)$.

The matrix $\mD$ has size $(d(d+2r) \times dr$; no more computations
are needed to fill its entries.
The matrix $\mD'$ has size $d(r+1) \times d(d+2r)$. Its entries are
obtained by evaluating $c_0,\dots,c_{r+d}$ at all $n$ in $B$. Since
$\deg(c_i)\le r$ and $|B|=d(r+1)$, this takes time $O(\M(dr)\log(dr))$
per polynomial. Since $r \le d$, the total time is
$T_3=O(d^2\M(r)\log(r)) \in \bigOsoft(d^2r)$.


The matrix $\mE=\mD'\mD$ has size $d(r+1) \times dr$; using block
matrix multiplication with blocks of size $dr$, it can be computed in
time $T_4=O(d^{\omega+1}r^{\omega-1})$. A basis of its nullspace can
be computed in time $T_5=O(d^\omega r^\omega)$.

Given a vector $[u_n \ | \ n\in A]$ in the nullspace of $\mE$, one can
reconstruct $[u_n \ | \ 0 \le n < pd]$ using~\eqref{eq:rec}. This
first requires evaluating all coefficients of all equations $C_{n-r}$,
for $n$ in $A'=[0,\dots,pd-1]-A$, which takes time $T_6=O(d 
\M(s)\log(s))$, with $s=\max(r,pd)$. 

Then, for a given $[u_n \ | \ n\in A]$ in the nullspace of $\mE$,
deducing $[u_n \ | \ 0 \le n < pd]$ requires $|A'|< pd$ matrix-vector
products in size $d+r$. The dimension of the nullspace is $O(dr)$; we
process all vectors in the nullspace basis simultaneously, so that we
are left to do $pd$ matrix products in size $(d+r) \times (d+r)$ by
$(d+r) \times dr$. The cost of each product is $O(d^\omega r)$, so the
total cost is $T_7=O(d^{\omega+1} r p)$.

Summing $T_1,\dots,T_5$ proves the first part of the
proposition. Adding to this $T_6$ and $T_7$ gives the second claim.


\subsection{Proof of Theorem~\ref{coro:solpol}}
 
\noindent Let $\mathcal{F}$ and $\mathcal{G}$ be as above. By
Proposition~\ref{theo:theorysolpol}, $\dim_{\F_p}(\mathcal{G})=0$ if and
only if $\dim_{\F_p(x^p)}(\mathcal{F})=0$. Hence, the first estimate
of Proposition~\ref{theo:solpol} proves our first claim.

Suppose that $\dim_{\F_p}(\mathcal{G})\ne 0$, and let $u_1,\dots,u_k$
be an $\F_p$-basis of $\mathcal{G}$.
Proposition~\ref{theo:theorysolpol} implies that $u_1,\dots,u_k$
generates $\mathcal{F}$ over $\F_p(x^p)$.  We deduce an $\F_p(x^p)$-basis
$\mathscr{B}$ of $\mathcal{F}$ in a naive way: starting from
$\mathscr{B}=[u_1]$, we successively try to add $u_2,\dots$ to
$\mathscr{B}$. Independence tests are performed at each step, using
the following lemma.

\begin{Lemma}
  Given $u_1,\dots,u_\ell$ in $\F_p[x]$ of degree less than $pd$, one
  can determine whether they are linearly independent over $\F_p(x^p)$
  in time $\bigOsoft(\ell^{\omega+2} dp)$.
\end{Lemma}
\myproof It suffices to compute their Wronskian determinant. The
determinant of a matrix of size $\ell$ can be computed using
$O(\ell^{\omega+1})$ sums and products~\cite{Berkowitz84}; since here
all products can be truncated in degree $\ell dp$, the cost is
$O(\ell^{\omega+1} \M(\ell dp))$. \foorp

\smallskip\noindent
At all times, there are at most $r$ elements in $\mathscr{B}$, so we
always have $\ell \le r+1$. Since we also have $k \le dr$, the overall
time is $\bigOsoft(d^2 r^{\omega+3} p)$, as claimed.



\section{p-curvature: first order}\label{sec:one}

\noindent For first order operators, there is a closed form formula
for the $p$-curvature. Let $L=\partial-u$, with $u$ in $\F_p(x)$;
then, by~\cite[Lemma 1.4.2]{vanDerPut95}, the $p$-curvature of $L$ is the
$1\times 1$ matrix with entry $u^{(p-1)}+u^p$, where the first term is
the derivative of order $p-1$ of $u$. In this case, we do not
distinguish between the $p$-curvature and its unique entry.

The case of first order operators stands out as the only one where a
cost polynomial in $\log(p)$ can be reached; this is possible since in
this case, we only compute $O(d)$ non-zero coefficients. As per our
convention, in the following statement, we take $L$ not necessarily
monic, but with polynomial coefficients.

\begin{theorem}\label{theo:ordre1}
  Given $L=a\partial -b$ in $\F_p[x]\langle \partial \rangle$ of
  bidegree $(d,1)$ that satisfies $\bf H$, one can compute its
  $p$-curvature in time $O(d\M(d)\log(p))\subset \bigOsoft(d^2
  \log(p))$.
\end{theorem}
\myproof Since the $p$-curvature belongs to $\F_p(x^p)$, it suffices
to compute its $p$th root. Computing the $p$-curvature itself requires
no extra arithmetic operation, since taking $p$-powers is free 
over $\F_p$, as far as arithmetic operations are concerned.
Hence, we claim that the rational function
$$\left (  \Big(\frac ba\Big )^{(p-1)} + \Big(\frac ba\Big )^p\right
)^{\frac 1p}
=\left (  \Big(\frac ba\Big )^{(p-1)} \right
)^{\frac 1p} + \frac ba
$$
can be computed in time $O(d\M(d)\log(p))$.  Of course, the only
non-trivial point is to compute $s=(u^{(p-1)})^{1/p}$, with $u=b/a$.

Observe that $a^p u^{(p-1)}$ is a polynomial of degree less than $dp$,
so $as$ is a polynomial of degree less than $d$. Hence, it is enough
to compute the power series expansion $s \bmod x^d$. From this, we 
deduce the polynomial $as$ by a power series multiplication in degree
$d$, and finally $s$ by division by $a$.

Let us write the power series expansion $u =\Sigma_{i \ge 0} u_i
x^i$. Then, the series $s$ equals $-\Sigma_{i \ge 0} u_{ip}x^i$, so it
is enough to compute the coefficients $(u_{ip})_{i < d}$.

We start by computing the first coefficients $u_0,\dots,u_{d-1}$ by
power series division, in time $O(\M(d))$. From these initial
conditions, the coefficients $u_p,\dots,u_{p+d-1}$ can be deduced for
$O(\M(d)\log(p))$ operations using binary powering techniques,
see~\cite{Fiduccia85} or~\cite[Sect.~3.3.3]{Bostan03}.  Iterating this
process $d$ times, we obtain the values $u_{ip},\dots,u_{ip+d-1}$, for
$i < d$, in time $O(d\M(d)\log(p))$.\foorp

\smallskip\noindent As an aside, note that by
Lemma~\ref{lemma:pcurv-vs-ratsols}, a rational function $u$ is a
logarithmic derivative in $\F_p(x)$ if and only if $u^{(p-1)} +
u^p=0$.  This point also forms the basis of Niederreiter's algorithm
for polynomial factoring~\cite{Niederreiter93}.


\section{p-curvature: second order}\label{sec:two}

\noindent For second order operators, it is possible to exploit a
certain linear differential system satisfied by the entries of the
$p$-curvature matrix: already in~\cite{Dwork90,vanDerPut96}, one finds
a third order linear differential equation satisfied by an
anti-diagonal entry of the $p$-curvature, for the case of operators of
the form $\partial^2+s$, or more generally $\partial^2+r\partial +s$,
when $r^{(p-1)}+r^p=0$.

In this section, we let $L$ have the form $v\partial^2+w\partial+u$,
with $u,v,w$ in $\F_p[x]$ of degree at most $d$. We assume that $d
\geq 2$ and $p >2$, and that $\bf H$ holds (we do not repeat these
assumptions in the theorems); we let $\mA$ be the companion matrix of
$L$ and let $\mA_p$ be its $p$-curvature.

We give partial results regarding the computation of $\mA_p$: we give
algorithms of cost $\bigOsoft(p^{1/2})$ or $\bigOsoft(p)$ to test
properties of $\mA_p$, or compute it in some cases, up maybe to some
indeterminacy.  Though these algorithms do not solve all questions,
they are still substantially faster than the ones for the general case
in the next section.

\smallskip\noindent{\bf The trace of the $p$-curvature.} We start by
an easy but useful consequence of the result of the previous
section: the trace of $\mA_p$ can be computed fast.
\begin{theorem}\label{theo:trace}
  One can compute the trace $\tau$ of $\mA_p$ in time
  $O(d\,\M(d)\log(p))$.
\end{theorem}
\myproof The $p$-curvature of a determinant connection is the trace of the
$p$-curvature of the original connection~\cite{Katz82,Voloch00}.  
Concretely, this
means that the trace of $\mA_p$ is equal to the $p$-curvature of $v\partial+w$.
By Theorem~\ref{theo:ordre1}, it can be computed in time
$O(d\,\M(d)\log(p))$.  \foorp

\smallskip\noindent{\bf Testing nilpotence.} As a consequence of the
previous theorems, we obtain a decision procedure for nilpotence.
\begin{Coro}\label{coro:nilpotence}
  One can decide whether $\mA_p$ is nilpotent in time
  $\bigOsoft(d^\omega p^{1/2}+ d^{\omega+1}).$
\end{Coro} 
\myproof The $p$-curvature $\mA_p$ is nilpotent if and only if its
trace and determinant are both zero. By Theorem~\ref{theo:trace}, the
condition on the trace can be checked in time logarithmic in~$p$. By
Lemma~\ref{lemma:pcurv-vs-ratsols}, the second condition
$\det(\mA_p)=0$ is equivalent to the fact that $Lu=0$ has a non-zero
solution, which can be tested in the requested time by
Theorem~\ref{coro:solpol}.  \foorp

\smallskip\noindent{\bf The eigenring.} To state our further results,
we need an extra object: the {\em eigenring} $\mathcal{E}(L)$ of
$L$. This is the set of matrices $\mB$ in $\mathscr{M}_2(\F_p(x))$
that satisfy the matrix differential equation
\begin{equation}
  \label{eq:eigenring}
  \mB' =  \mB \mA - \mA \mB
\end{equation}
(our definition differs slightly from the usual one in the sign
convention). By construction, the eigenring $\mathcal{E}(L)$ is a
$\F_p(x^p)$-vector space of dimension at most 4, which contains the
$p$-curvature $\mA_p$. Then, we let $\gamma$ be its dimension over
$\F_p(x^p)$; we will prove later on that $\gamma$ is in $\{2,4\}$.

Let further $\mathcal{F}$ be the set of solutions of $Ly=0$ in
$\F_p(x)$ and let $\beta$ be its dimension over $\F_p(x^p)$.  Then,
our main results are the following.
\begin{theorem}\label{theo:main2}
  One can compute in time $\bigOsoft(d^{\omega+1}p):$
  \begin{enumerate}
  \item[1.] the dimensions $\gamma \!\in\! \{2,4 \}$ of $\mathcal{E}(L)$
    and $\beta \!\in\! \{0,1,2 \}$~of~$\mathcal{F}$;
  \item[2.] $\mA_p$, if $\gamma=4$ or $\beta=2$.
  \item[3.] $\mA_p$, up to a multiplicative constant in $\F_p[x^p]$ of
    degree at most $pd$, if $\gamma=2$ and the trace $\tau=0$.
  \item[4.] a list of two candidates for $\mA_p$, if $\gamma=2$ and
    $\beta=1$.
\end{enumerate}        
\end{theorem}
\noindent The rest of this section is devoted to prove this theorem.

\smallskip\noindent{\bf The dimension of the eigenring.} The following
lemmas restrict the possible dimension $\gamma$ of $\mathcal{E}(L)$.

\begin{Lemma}\label{lemma:4.a}
  If $\mA_p$ has the form $\lambda \mI_2$, then $\gamma=4$.
\end{Lemma}
\myproof In this case, the commutator of $\mA_p$ in
$\mathscr{M}_2(\F_p(x))$ is $\mathscr{M}_2(\F_p(x))$ itself, so it has
dimension 4 over $\F_p(x)$. Then,~\cite[Prop.~3.5]{Cluzeau04} implies
that $\mathcal{E}(L)$ has dimension~4 over $\F_p(x^p)$. \foorp
 
\begin{Lemma}\label{lemma:4}
  Either $\gamma=2$, or $\gamma=4$. In the second case, $\mA_p$ is
  equal to $\frac{\tau}{2}\, \mI_2$, where $\tau$ is the trace of
  $\mA_p$.
\end{Lemma}
\myproof Corollary~1 of~\cite{Cluzeau04} shows that if the minimal and
characteristic polynomials of $\mA_p$ coincide, then $\mathcal{E}(L)$
equals $\F_p(x^p)[\mA_p]$. In this case, $\F_p(x^p)[\mA_p]$ has
dimension 2 over $\F_p(x^p)$.  Else, the minimal polynomial of $\mA_p$
must have degree 1, so $\mA_p$ is necessarily equal to
$\frac{\tau}{2}\, \mI_2$, and we are under the assumptions of the previous
lemma. \foorp

\smallskip\noindent{\bf Computing $\gamma$ and $\beta$.}
The equality~\eqref{eq:eigenring} gives a system of four linear
differential equations of order one for the entries
$b_{1,1},\dots,b_{2,2}$ of $\mB$. An easy computation shows
that~\eqref{eq:eigenring} is equivalent to the system
\begin{eqnarray}\label{eq:3}
v^3 b_{2,1}''' + A b_{2,1}' + B b_{2,1} &=& 0,\\ \label{eq:4} 
v^2 b_{1,2} + R b_{2,1}'' + S b_{2,1}' + T b_{2,1} &=& 0,\\  \label{eq:5}
v(b_{1,1}-b_{2,2}) + v b_{2,1}' -w b_{2,1} &=& 0,\\\label{eq:6}
b_{1,1}'+b_{2,2}' &=& 0, \label{eq:7} 
\end{eqnarray}
where $A,B,R,S,T$ belong to $\F_p[x]$, and are given by 
$$A=v(-2{w'}v+2w{v'}+4uv-w^2),$$
$$B=vw(v''-w')+
{v'}w(w-2v')+2u'v^2-2vu{v'}-{w''}v^2+2{v'}{w'}v$$ and
$$R = v^2/2, \quad S = -vw/2, \quad T = v' w/2 - v w'/2 + uv.$$
Since Equation~\eqref{eq:6} is equivalent to $b_{1,1}+b_{2,2} \in
\F_p(x^p)$, we readily deduce that the dimension $\gamma$ 
of $\mathcal{E}(L)$ equals $\gamma'+1$, where $\gamma'$
is the dimension of the solution-set of~\eqref{eq:3}.

Computing both $\gamma$ and $\beta$ can be done using
Theorem~\ref{coro:solpol}, with respectively $r=3$ or $r=2$, and in
degree respectively at most $4d$ or $d$. This proves point~1 of
Theorem~\ref{theo:main2}.

If $\gamma=4$, we are in the second case of Lemma~\ref{lemma:4}. Since
the trace can be computed in time $\bigOsoft(d^2\log(p))$ by
Theorem~\ref{theo:trace}, point~2 of Theorem~\ref{theo:main2} is
established in this case. If $\beta=2$, then $\mA_p$ is zero by
Lemma~\ref{lemma:pcurv-vs-ratsols}, so point~2 of
Theorem~\ref{theo:main2} is established as well.

\smallskip\noindent{\bf Eigenrings of dimension $2$.}  The rest of
this section is devoted to analyze what happens if $\mathcal{E}(L)$
has dimension $\gamma=2$ over $\F_p(x^p)$, so that the dimension
$\gamma'$ of the solution-space of~\eqref{eq:3} is 1. In this case,
the information provided by the eigenring is not sufficient to
completely determine the $p$-curvature. However, it is still possible
to recover some useful partial information. To fix notation,
we write the $p$-curvature as 
$$\mA_p = \left [ \begin{matrix}
  f_{1,1} & f_{1,2}\\
  f_{2,1} & f_{2,2}
\end{matrix} \right ].$$

\begin{Lemma}\label{lemma:dim1}
  If $\gamma=2$, $F=v^p f_{2,1}$ is a nonzero polynomial solution of
  degree at most~$pd$ of Equation~\eqref{eq:3}.
\end{Lemma}
\myproof Since the $p$-curvature $\mA_p$ belongs to the eigenring, its
entries $f_{1,1},\dots,f_{2,2}$ satisfy~\eqref{eq:3}
to~\eqref{eq:6}. Lemma~\ref{lemma:1} shows that $F=v^p f_{2,1}$ is a
polynomial solution of degree at most~$pd$ of Equation~\eqref{eq:3}.
Moreover, $F$ cannot be $0$, since otherwise Equations~\eqref{eq:4}
to~\eqref{eq:6} would imply that $\mA_p$ has the form $\lambda \mI_2$
for some $\lambda$ in $\F_p(x^p)$. By Lemma~\ref{lemma:4.a}, this
would contradict the assumption $\gamma=2$ .  \foorp

\begin{Lemma}\label{lemma:dim2}
  Suppose that $\gamma=2$ and let $u \in \F_p[x]$ be the nontrivial
  polynomial solution of minimal degree of
  Equation~\eqref{eq:3}. There exists a nonzero polynomial $c$ in
  $\F_p[x^p]$ of degree at most~$pd$, such that the entries of $\mA_p$
  are given by
\begin{eqnarray*}
f_{1,1} &=& \frac12 \left( \tau + \frac{c}{v^p} \left(\frac{w}{v} u-u' \right) \right),\\ 
f_{1,2} &=& - \frac{c}{v^{p+2}} \left(R u'' + Su' + Tu \right),\\ 
f_{2,1} &=& \frac{c}{v^p} u,\\
f_{2,2} &=& \frac12 \left( \tau - \frac{c}{v^p} \left(\frac{w}{v} u-u' \right) \right).
\end{eqnarray*}
\end{Lemma}
\myproof By Lemma~\ref{lemma:dim1}, the polynomials $F$ and $u$ both
satisfy Equation~\eqref{eq:3}; thus, they differ by an element $c$ in
$\F_p(x^p)$. Moreover, the minimality of the degree of $u$ implies
that $c=F/u$ actually belongs to $\F_p[x^p]$ and has degree at most
$\deg(F) \leq pd$.  The rest of the assertion follows from the
relations $F=v^p f_{2,1}$, $\tau = f_{1,1} + f_{2,2}$ and the
equalities~\eqref{eq:4} and~\eqref{eq:5}. \foorp

\smallskip\noindent{\bf Concluding the proof of Theorem~\ref{theo:main2}.}
To conclude, we consider two special cases. If $\tau=0$, as
in~\cite{vanDerPut96}, the previous lemma shows that $\mA_p$ is known
up to a multiplicative constant in $\F_p[x^p]$, as soon as the
polynomial $u$ has been computed. In this case,
Corollary~\ref{coro:solpol} shows that one can compute a non-zero
solution $u_0$ of~\eqref{eq:3} in the required time. The minimal
degree solution $u$ by clearing out the factor in $\F_p[x^p]$ in $u_0$
using~\cite[Ex.~14.27]{GaGe99}, in negligible time $O(\M(dp)\log(dp))
\subset \bigOsoft(dp)$, and the substitution in the former formulas
takes time $\bigOsoft(dp)$ as well.

If $\beta=1$, $L$ has a non-trivial polynomial solution, so by
Lemma~\ref{lemma:pcurv-vs-ratsols} the determinant of $\mA_p$ is zero;
the additional equation $f_{1,1} f_{2,2}=f_{1,2}f_{2,1}$, in
conjunction with the formulas in Proposition~\ref{lemma:dim1},
uniquely determines the polynomial $c^2$ and thus leaves us with only
two possible candidates for $\mA_p$.



\section{p-curvature: higher order}\label{sec:high}

\noindent In this final section, we study operators of higher order,
and we prove that the $p$-curvature can be computed in time
subquadratic in $p$. 
\begin{theorem}
  Given $L$ in $\F_p[x]\langle \partial \rangle$ of bidegree $(d,r)$,
  one can compute its $p$-curvature in time
$$\bigOsoft(r^\omega d^2 p^{2\omega/3} + r^\omega d p^{1+\omega/3}).$$
\end{theorem}
Hence, the exponent in $p$ is ${1+\omega/3} < 1.79 < 2$; in the best
possible case $\omega=2$, we would obtain an exponent ${5/3}$ in $p$,
unfortunately still not optimal.

As a result of independent interest, we also give an algorithm for
computing the image of a matrix of rational functions by an
differential operator similar in spirit to Brent and Kung's algorithm
for modular composition~\cite{BrKu78}; to our knowledge, no prior
non-trivial algorithm existed for this task.


\subsection{Preliminaries}

\smallskip\noindent{\bf Euler's operator.}
Besides operators in the usual variables $x,\partial$, it will also be
convenient to consider operators in $\F_p(x)\langle \theta \rangle$ or
$\mathscr{M}_n(\F_p(x))\langle \theta \rangle$, where $\theta$ is
Euler's operator~$x \partial$, which satisfies the commutation rule
$\theta x = x\theta + x$. To avoid confusion, we may say that $L$ has
bidegree $(d,r)$ {\em in $\partial$} or {\em in $\theta$}, if $L$ is
written respectively on the bases $(x,\partial)$ or $(x,\theta)$.

\smallskip\noindent{\bf Conversion.}
Given an operator $L$ in $\mathscr{M}_n(\F_p[x])\langle \partial
\rangle$ of bidegree $(d,r)$, $L'=x^r L$ can be rewritten as an
operator in $\theta$ with polynomial coefficients. The operator $L'$
has bidegree $(d+r,r)$ in $\theta$. By~\cite[Section~3.3]{BoChLe08},
computing the coefficients of $L'$ takes time $O(n^2
(d+r)\M(r)\log(r))$. Since representing all coefficients of $L'$
requires $O(n^2 (d+r)r)$ elements, this is quasi-linear, up to
logarithmic factors.

\smallskip\noindent{\bf Multiplication.} Next, we give an algorithm
for the multiplication of operators with rational coefficients of a
special type, inspired by that of~\cite{BoChLe08} (which handles
polynomial coefficients). The algorithm relies on an evaluation /
interpolation idea originally due to~\cite{vanDerHoeven02}, and 
introduces fast matrix multiplication to solve the problem.

\begin{Lemma}\label{lemma:mul}
  Let $b\in \F_p[x]$ be of degree at most $d$, with $b(0)\ne 0$ and
  let $\gamma,\mu$ be in $\F_p(x)\langle\partial \rangle$, with
  $$\gamma=\sum_{j=0}^{h} \frac{g_{j}}{b^{h-j}} \partial^j,\quad 
  \mu=\sum_{j=0}^{h} \frac{m_{j}}{b^{h-j}} \partial^j,$$ where $g_j$
  and $m_j\in \F_p[x]$ have degrees at most $d(h-j)$. Then if $2h \le
  p-1$, one can compute $\eta=\gamma \mu$ in time $O(h^\omega d^2)$.
\end{Lemma}
\myproof Define $\eta^\star = b^{2h} \eta,\gamma^\star = b^{h} \gamma$
and $\mu^\star = b^{h} \mu$. A quick verification shows that these
operators are in $\F_p[x]\langle \partial \rangle$, of respective
bidegrees bounded by $(2dh,2h)$, $(dh,h)$ and $(dh,h)$.

We first compute $\gamma(x^j)$ and $\mu(x^j) \bmod x^{2dh+h+1}$ for $j
\le 2h$. This is done by computing the corresponding values of
$\gamma^\star$ and $\mu^\star$, and dividing the results by $b^h$. The
former computation takes time $O(\M(dh^2))$ using algorithm {\sf Eval}
of~\cite{BoChLe08}; the latter $O(h\M(dh))$ by Newton iteration for
power series division. Our assumption $2h \le p-1$ ensures that
divisions performed in the evaluation algorithm (and in the
interpolation below) are well-defined.

From the values of $\gamma$ and $\mu$, the values $\eta(x^j)
\bmod x^{2dh+1}$ are obtained as in~\cite[Th. 3]{BoChLe08}; the cost is
$O(h^\omega d^2)$. We can then compute the values of $\eta^\star$ in
time $O(h\M(dh))$ by fast polynomial multiplication.  Knowing its
values, we recover $\eta^\star$ using algorithm {\sf Interpol}
of~\cite{BoChLe08}; this takes time $O(\M(dh^2))$. Finally, we deduce
$\eta$ by division by $b^{2h}$; this takes time $O(h\M(dh)\log(dh))$,
using fast gcd computation.  \foorp


\smallskip\noindent{\bf Left and right forms.} Let $L\in
\mathscr{M}_n(\F_p[x])\langle \theta\rangle$ have the form
$$L = \bell_0(x) + \bell_1(x) \theta + \cdots + \bell_r(x) \theta^r,$$
with $\bell_i \in \mathscr{M}_n(\F_p[x])$ of degrees at most $d$. It
can be rewritten
$$L = \bell^\star_0(x) +  \theta^\star \bell_1(x) + \cdots + \theta^r \bell^\star_r(x),$$
with $\bell^\star_i$ in $\mathscr{M}_n(\F_p[x])$ of degrees at most $d$
as well. The former expression will be called the {\em right-form} of
$L$; the latter is its {\em left-form}. 

\begin{Lemma}\label{lemma:leftright}
  Let $L$ have bidegree $(d,r)$ in $\mathscr{M}_n(\F_p[x])\langle
  \theta\rangle$, given in its right-form (resp. in its left-form). Then
  \sloppy one can compute its left-form (resp. right-form) in time
  $O(n^2d \M(r)\log(r))\subset \bigOsoft(n^2 dr)$.
\end{Lemma}
\myproof We prove one direction only; the other is similar. Given the
right-form of $L$, we can (without performing any operation) rewrite
$\begin{array}{c}L=\sum_{j \le d} x^j L_i,\end{array}$ where $L_j$ has
constant coefficients and order at most $r$.  Since $x^j L_i(\theta) =
L_i(\theta-j)x^j$, the result follows by using algorithms for
polynomial shift by $j$~\cite{GaGe97}.  \foorp

\smallskip\noindent The number of elements needed to represent $L$ in
either left- or right-form is $O(n^2dr)$, so the previous algorithm is
quasi-linear, up to logarithmic factors.


\subsection{Evaluation}

\noindent For $L$ in $\mathscr{M}_r(\F_p[x])\langle \partial
\rangle$ or $\mathscr{M}_r(\F_p[x])\langle \theta \rangle$ and $\mA$
in $\mathscr{M}_r(\F_p(x))$, $L\mA$ denotes the matrix in
$\mathscr{M}_r(\F_p(x))$ obtained by applying $L$ to $\mA$. In this
subsection, we give cost estimates on the computation of $L\mA$.

\smallskip\noindent{\bf The polynomial case.} We start with the case
of an operator with polynomial coefficients, which we apply to a
matrix with polynomial entries. We use an operator in $\theta$, since
this makes operations slightly more convenient than in $\partial$. As
in Section~\ref{sec:solpol}, we make assumptions on the relative sizes
of the input parameters (here $\delta, \rho, \varepsilon$), for
simplicity's sake. 
\begin{Lemma}\label{Prop:2}
  Given $L \in \mathscr{M}_r(\F_p[x])\langle \theta \rangle$ of
  bidegree $(\delta,\rho)$ and $\mE \in \mathscr{M}_r(\F_p[x])$ of
  degree $\varepsilon$, one can compute $L \mE$ in time
  $\bigOsoft(r^\omega \rho \varepsilon^{\omega-2} \delta^{3-\omega})$,
  assuming $\delta \in O(\varepsilon)$ and $\varepsilon \in
  O(\rho^{1/2} \delta)$.
\end{Lemma}
The cost can be rewritten as $\bigOsoft(r^\omega \rho\, \varepsilon\,
(\delta/\varepsilon)^{3-\omega})$. Since $\omega \le 3$ and $\delta \in O(\varepsilon)$, this is always better than $\bigOsoft(r^\omega \rho
\varepsilon)$: the cost ranges from $\bigOsoft(r^\omega \rho \delta)$
for a hypothetical $\omega=2$ to $\bigOsoft(r^\omega \rho
\varepsilon)$ for $\omega=3$. As a matter of comparison, let us write
$$
\begin{array}{c}
L=\sum_{i\le \rho}\bell_i \theta^i, \quad \bell_i \in \mathscr{M}_r(\F_p[x]).
\end{array}
$$Computing $L\mE$
naively amounts to computing all $\theta^i \mE$ for $i \le \rho$,
multiplying them by the respective coefficients $\bell_i$, and summing
the results; the cost is in $\bigOsoft(r^\omega \rho \varepsilon),$ so
our estimate is better. 

\smallskip\noindent\myproof Our result is achieved using a baby
steps~/~giant steps strategy inspired by Brent-Kung's algorithm for
power series composition~\cite{BrKu78}. Let $k=\lfloor
\rho^{1/2}\rfloor$ and $h=\lceil \rho/k\rceil$. First, we rewrite $L$
in left-form, as
$$\begin{array}{c}L=\sum_{i \le \rho} \theta^i\bell^\star_i(x);\end{array}$$ 
by Lemma~\ref{lemma:leftright}, the cost is $T_1\!=\!O(r^2 \delta
\M(\rho)\log(\rho))\subset \bigOsoft(r^2 \delta \rho)$.
Next, $L$ is cut into $h$ slices of the form
$$
\begin{array}{c}
L_0 + \theta^k L_1 +  \cdots + \theta^{(h-1)k} L_{h-1},
\quad\text{\it i.e.}\quad
L=\sum_{j < h} \theta^{jk}L_j.
\end{array}
$$
Each $L_j$ has order less than $k$ and can be written as
$$\begin{array}{c}L_j = \sum_{i < k} \theta^i \bell^\star_{jk+i}(x),\end{array}$$ 
where for $jk+i > \rho$,  $\bell^\star_{jk+i}$ is
zero. Finally, we rewrite each $L_j$ in right-form:
\begin{equation}\label{eq:L''}
\begin{array}{c}
L_j = \sum_{i < k}  \bell^\dagger_{j,i}(x) \theta^i,  
\end{array}
\end{equation}
where all $\bell^\dagger_{j,i}$ have degree at most $\delta$. By
Lemma~\ref{lemma:leftright}, the cost is $T_2=O(hr^2\delta
\M(k)\log(k))$, which is in $\bigOsoft(r^2\delta \rho)$ as before.  
To apply
$L$ to $\mE$, we first compute the baby steps
$$\mE_0=\mE,\ \mE_1=\theta \mE,\ \dots,\ \mE_{k-1}=\theta^{k-1} \mE;$$
then, we deduce all $L_j \mE$, for $j<h$; finally, we do the giant
steps
$$\begin{array}{c}L \mE=\sum_{j<h} \theta^{jk} L_j \mE.\end{array}$$
All $\mE_i$ can be computed in time $T_3=O(r^2 \rho^{1/2}\varepsilon
)$, by successive applications of $\theta$. The cost $T_4$ of deducing
the polynomials $L_j \mE$ is detailed below. Finally, one recovers $L
\mE$ by first computing all $\theta^{jk} L_j \mE$, for $j<h$, and then
summing them. Since $\theta^i(x^j) = j^i x^j$, $\theta^{jk}$ can be
applied to $L_j \mE$ in time $O(r^2\varepsilon\log(\rho))$, so the
total cost of this final step is $T_5=O(r^2 \varepsilon h
\log(\rho))\subset \bigOsoft(r^2 \rho^{1/2} \varepsilon)$.

It remains to compute all $L_j \mE$, given all $\mE_i$; we compute
them all at once. In view of Equation~\eqref{eq:L''}, we have
$$\begin{array}{c}L_j \mE=\sum_{i < k}  \bell^\dagger_{j,i} \mE_i,\end{array}$$
where the $\mE_i$ are known. We cut $\mE_i$ into slices of length $\delta$:
$$\begin{array}{c}\mE_i = \sum_{u < s} \mE_{i,u} x^{\delta u},\end{array}$$ 
where $\mE_{i,u}$ has degree less than $\delta$ and $s =\lceil
\varepsilon/\delta \rceil \le 2\varepsilon/\delta$. This gives
$$L_j \mE = \sum_{i < k}  \bell^\dagger_{j,i} \sum_{u < s} \mE_{i,u}x^{\delta u}
= \sum_{u < s} x^{\delta u}  \sum_{i < k}\bell^\dagger_{j,i} \mE_{i,u}.$$
We will compute all inner sums
$$\begin{array}{c}\sum_{i < k} \bell^\dagger_{j,i} \mE_{i,u}\end{array}$$
at once, for $j < h$ and $u < s$; from this, one can recover all
$L_j\mE$ in time $O(r^2 \varepsilon \rho^{1/2})$.

The computation of these sums amounts to perform a $(h\times k) \times
(k\times s)$ matrix multiplication, with entries that are polynomial
matrices of size $r$ and degree at most $\delta$.  Since $\varepsilon
\in O(\rho^{1/2}\delta)$, we have $s \in O(\rho^{1/2})$. Hence, we
divide the previous matrices into blocks of size $s$ and we are left
to do a $(O(\rho^{1/2}/s) \times O(\rho^{1/2}/s)) \times
(O(\rho^{1/2}/s)\times O(1))$ product of such blocks, where
$\rho^{1/2}/s$ is lower-bounded by a constant.  Multiplying a single
block takes time $O(r^\omega s^\omega \M(\delta))$, so the total time
$T_4$ is $O(r^\omega \rho s^{\omega-2}\M(\delta))$, which is 
$\bigOsoft(r^\omega \rho \varepsilon^{\omega-2} \delta^{3-\omega})$.

The conclusion of Lemma~\ref{Prop:2} comes after a few
simplifications, which shows that the dominant cost is $T_4$, for the
final linear algebra step. \foorp

\smallskip\noindent{\bf The rational function case.} Next, we study
the application of an operator to a matrix of rational functions~$\mA$
(we make some simplifying assumptions on the denominators in~$\mA$,
which will be satisfied in the cases in \S 6.3 where we apply this
result). Besides, our operator is now in
$\mathscr{M}_r(\F_p[x])\langle \partial \rangle$ rather than in
$\mathscr{M}_r(\F_p[x])\langle \theta \rangle$.

Because of the larger number of parameters appearing in the
construction, the cost estimate unfortunately becomes more complex than
in the polynomial case.
\begin{Lemma}\label{prop:eval}
  Let $L \in \mathscr{M}_r(\F_p[x])\langle \partial \rangle$ be of
  bidegree $(\delta,\rho)$. Let $\mA \in \mathscr{M}_r(\F_p(x))$ be of
  the form $\mB/b^\kappa$, with $b \in \F_p[x]$ of degree at most $d$
  and $\mB \in \mathscr{M}_r(\F_p[x])$ of degree at most $\kappa
  d$. Define
  $$\delta'=\delta+\rho\quad\text{and}\quad\varepsilon =
  (\kappa+\rho)d+\delta'+1.$$ If $b(0)\ne 0$ and $\varepsilon \in
  O(\rho^{1/2} \delta')$, one can compute $L \mA$ in time 
  $\bigOsoft(r^\omega \rho \varepsilon^{\omega-2} {\delta'}^{3-\omega})$.
\end{Lemma}
\myproof Let $L'=x^\rho L$. Given $L$ as an operator in $\partial$, we
saw that we can write $L'$ as an operator in $\theta$, of bidegree
$(\delta',\rho)$; the coefficients of $L'$ in $\theta$ can be computed
in time $O(r^2 \delta \M(\rho)\log(\rho)) \subset \bigOsoft(r^2 \delta
\rho)$. To conclude, it is enough to compute $L' \mA$, since then $L
\mA$ is deduced by a division by $x^\rho$, which is free.

For any $i\ge 0$, $\theta^i \mA$ has the form $\mB_i/b^{\kappa+i}$,
with $\mB_i$ in $\mathscr{M}_r(\F_p[x])$ of degree at most
$(\kappa+i)d$. Thus, $L' \mA$ has the form
$\mB^\star/b^{\kappa+\rho}$, with $\mB^\star \in
\mathscr{M}_r(\F_p[x])$ of degree less than $\varepsilon$, with
$\varepsilon=(\kappa+\rho)d+\delta'+1$.

Knowing $L' \mA \bmod x^\varepsilon$, one can recover the numerator
matrix $\mB^\star$ through multiplication by $b^{\kappa+\rho}$; a gcd
computation finally gives $L'\mA$ in normal form. These latter steps
take time $O(r^2 \M(\varepsilon)\log(\varepsilon))\subset
\bigOsoft(r^2 \varepsilon)$.

Since $b(0)\ne 0$, the matrix $\mE=\mA \bmod x^\varepsilon$ is
well-defined; it can be computed in time $O(r^2
\M(\varepsilon))\subset \bigOsoft(r^2 \varepsilon)$ by power series
division. Lemma~\ref{Prop:2} gives complexity estimates for computing
$L'\mE \bmod x^\varepsilon$. Since this matrix coincides with $L'\mA$
modulo $x^\varepsilon$, this concludes the proof of the lemma, as all
previous costs are negligible compared to the one of
Lemma~\ref{Prop:2}. \foorp


\subsection{Computing the p-curvature}
\noindent Let $L$ be in $\F_p[x]\langle \partial \rangle$ of bidegree
$(d,r)$ and let $\mA$ be its companion matrix. We define the operator
$\Lambda \in \mathscr{M}_r(\F_p(x))\langle \partial \rangle$ as
$\Lambda = \partial + \mA;$ thus, as pointed out in the introduction,
the $p$-curvature of $L$ is obtained by applying $\Lambda^{p-1}$ to
$\mA$.

To obtain a cost better than $O(p^2)$, we first compute a high enough
power $\Lambda'=\Lambda^k$ of $\Lambda$; then, we apply $\Lambda'$ to
$\mA$ $k'$ times, with $k'\simeq (p-1)/k$. Since $p-1$ may not factor
exactly as $kk'$, a few iterations of this process are needed.

\smallskip\noindent{\bf Computing $\Lambda^k$.} Let $\ell=\ell_r \in
\F_p[x]$ be the leading coefficient of $L$. Then, $\Lambda$ has the
form $\partial+ \blambda/\ell$, with $\blambda$ in
$\mathscr{M}_r(\F_p[x])$ and $\ell \in \F_p[x]$ of degree at most $d$
and $\ell(0)\ne 0$. More generally, for $k \ge 0$, we can write
$\Lambda^k$ as
$$\begin{array}{c}\Lambda^k =\sum_{j=0}^k \blambda_{k,j} \partial^j,
\end{array} \quad 
\text{with}\quad \blambda_{k,j}=\frac{\bell_{k,j}}{b^{k-j}}$$
and $\bell_{k,j}$ in $\mathscr{M}_r(\F_p[x])$ of degree at most $d (k-j)$.  

\begin{Lemma}\label{lemma:powerK}
  If $k \le p-1$, one can compute $\Lambda^k$ in time $O(r^\omega k^\omega d^2)$.
\end{Lemma}
\myproof We use a divide-and-conquer scheme. Let $h=\lfloor k/2
\rfloor$; we assume for simplicity that $k=2h$; if $k$ is odd, an
extra (cheaper) multiplication by $\Lambda$ is needed. We assume that
$\Lambda^{h}$ is known, and we see it as an $r \times r$ matrix with
entries that are scalar operators; hence, to compute $\Lambda^k$, we
do $O(r^\omega)$ products of such scalar operators. All these products
have the form $\eta=\gamma \mu$ of the form seen in
Lemma~\ref{lemma:mul}, so each of their costs is $O(h^\omega d^2)=O(k^\omega d^2)$.
\foorp

\smallskip\noindent{\bf Computing $\Lambda^{kk'}\mA$.}  We fix $k\le p$,
and we compute the operators $\Gamma=\Lambda^k$ and $\Gamma'=d^k
\Gamma$. Writing $k'=\lfloor (p-1)/k \rfloor$,
we compute the sequence
$$\mA_{(1)} = \mA,\quad \mA_{(i)} = \Gamma \mA_{(i-1)},\quad
i=2,\dots,k',$$ so that $\mA_{(k')}=\Lambda^{kk'}\mA$. Thus, we have
$\mA_{(k')}=\mA_{kk'}$, where the latter matrix is defined in
Equation~\eqref{eq:pcurv}. Using the subroutines seen before, a quick
analysis not reproduced here shows that the optimal choice is
$k=\lfloor (p-1)^{2/3} \rfloor$. Then, computing $\Gamma$ takes time
$O(r^\omega k^\omega d^2)=O(r^\omega p^{2\omega/3} d^2)$ by
Lemma~\ref{lemma:powerK}.

By Lemma~\ref{lemma:1}, each matrix $\mA_{(i)}$ has the form
$\mB_{(i)}/b^{ik}$, with $\mB_{(i)}\in \mathscr{M}_r(\F_p[x])$ of
degree at most $dki$. Given $\mA_{(i)}$, we compute $\mA_{(i+1)}$ by
first applying $\Gamma'$ to $\mA_{(i)}$ and dividing the result by
$d^k$. 

The first step, applying $\Gamma'$, is the more costly. We obtain its
cost by applying Lemma~\ref{prop:eval}, with $(\delta,\rho)=(dk,k)$
and $\kappa=ik$. Then, we have $\delta'\in O(dk)$ and $\varepsilon \in
O(idk)$. For all $i\le k'$, we are under the assumptions of that
lemma; after a few simplifications, the cost becomes
$\bigOsoft(r^\omega i^{\omega-2} d k^2)$. Summing over all $i \le k'$,
we obtain an overall cost of $\bigOsoft(r^\omega {k'}^{\omega-1} d
k^2)$.  Taking into account that $k\in O(p^{2/3})$ and $k'\in
O(p^{1/3})$, this finally gives a cost of $\bigOsoft(r^\omega d
p^{1+\omega/3})$ for computing $\Lambda^{kk'}\mA$.

\smallskip\noindent{\bf Computing the $p$-curvature.} The definitions
of $k,k'$ imply that $p-(p-1)^{2/3} \le kk' \le p-1$. To obtain the
$p$-curvature $\Lambda^{p-1} \mA$, we iterate the previous process,
replacing the required number of steps $p-1$ by $p-1-kk'$, until the
required number of steps is $O(1)$. Since $p-1-kk' \le (p-1)^{2/3}$, it
takes $O(\log \log p)$ iterations; hence, the overall time is still in
$\bigOsoft(r^\omega d p^{1+\omega/3})$.

\smallskip\noindent{\bf Acknowledgments.} We wish to acknowledge financial
support from the French National Agency for Research (ANR Project ``Gecko"),
the joint Inria-Microsoft Research Centre, NSERC and the Canada Research Chair
program.

\smallskip
\bibliographystyle{abbrv}
\bibliography{curvature}

\end{document}